\documentclass{emulateapj}

\shorttitle{Velocity Dispersion in Dissipative Mergers}
\shortauthors{Stickley \& Canalizo}

\usepackage{natbib}
\usepackage{graphics}

\usepackage{color}

\newcommand{\sigmath}{\ensuremath \sigma_{\!*}}
\newcommand{\sig}{$\sigmath$}
\newcommand{\sigd}{$\sigma_{\rm d}$}
\newcommand{\msig}{\mbox{$M_{\rm BH}$--$\sigmath$}} 
\newcommand{\mbh}{$M_{\rm BH}$}
\newcommand{\smbh}{SMBH}
\newcommand{\smbhs}{SMBHs}
\newcommand{\gadget}{GADGET-3}
\newcommand{\GSnap}{\textsc{GSnap}}
\newcommand{\qeos}{\ensuremath q_{\rm eos}}
\newcommand{\sigo}{\ensuremath \sigma_0}
\newcommand{\kmps}{\ensuremath\rm {km~s^{-1}}}
\newcommand{\kmpst}{$\kmps$}
\newcommand{\ergps}{\ensuremath \rm erg\,s^{-1}}

\begin{document}

\title{STELLAR VELOCITY DISPERSION IN DISSIPATIVE GALAXY MERGERS WITH STAR FORMATION}

\author{\sc Nathaniel R. Stickley and Gabriela Canalizo}
\affil{Department of Physics and Astronomy, University of California,
    Riverside, CA 92521, USA}

\begin{abstract}

In order to better understand stellar dynamics in merging systems, such as NGC 6240, we examine the evolution of central
stellar velocity dispersion (\sig) in dissipative galaxy mergers using a suite of binary disk merger simulations that
include feedback from stellar formation and active galactic nuclei (AGNs). We find that \sig\ undergoes the same
general stages of evolution that were observed in our previous dissipationless simulations: coherent oscillation, then
phase mixing, followed by dynamical equilibrium. We also find that measurements of \sig\ that are based
only upon the youngest stars in simulations consistently yield lower values than measurements based upon the total
stellar population. This finding appears to be consistent with the so-called ``\sig\ discrepancy,''
observed in real galaxies. We note that quasar-level AGN activity is much more likely to occur when \sig\ is near its
equilibrium value rather than during periods of extreme \sig. Finally, we provide estimates of the scatter inherent in
measuring \sig\ in ongoing mergers.

\end{abstract}


\section{INTRODUCTION}

\noindent
Although measurements of \sig\ are often made in merging systems, such as NGC 6240
\citep{oliva1999,tecza2000,engel2010,medling2011}, little theoretical work has been done toward understanding the
detailed evolution of \sig\ during the merger process. Instead, most theoretical work involving \sig\ has focused on
passively evolving galaxy merger remnants. It is unclear whether \sig\ in a merging system is likely to be elevated or
suppressed compared with its fiducial, equilibrium value; the variability of \sig\ during the merger process is unknown.
The time required for \sig\ to reach a stable value is also unknown. These uncertainties impact any observational
program in which \sig\ is measured in potentially non-equilibrium systems. In particular, studies involving the \msig\
relation \citep{ferrarese2000,gebhardt2000, tremaine2002, gultekin2009, mcconnell2013} or the Fundamental Plane (FP) of
elliptical galaxies \citep{djorgovski1987,dressler1987,davies1987,bender1992} would benefit from a more complete
understanding of \sig\ in non-equilibrium systems.

The cosmological evolution of the \msig\ relation, a tight relationship between the mass of the central supermassive
black hole (\smbh) and \sig, may provide insights into the formation and growth histories of galaxies and \smbhs.
Several observational programs \citep[e.g.,][]{treu2004,treu2007,woo2006,woo2008,hiner2012,canalizo2012} that study the
cosmological evolution of the \msig\ relation include measurements of \sig\ in ongoing or recent mergers.
Unfortunately, the general lack of knowledge regarding the proper interpretation of \sig\ in such systems has cast
some doubt on the validity of using these systems to study of \msig\ evolution. For example, it is unknown whether
these systems have unusual velocity dispersions compared with systems that are clearly in a state of dynamic
equilibrium. Understanding the effect of measuring \sig\ in apparently non-relaxed systems would allow for a more
informed interpretation of these observations.

The FP is a relation among \sig, the half-light radius of a spheroid, and the mean surface brightness within the
half-light radius. It has been used to determine whether systems resemble normal elliptical galaxies \citep{woo2004,
rothberg2006}, but the FP is perhaps more useful as a tool for estimating distances to galaxies. Since \sig\ and
surface brightness are both independent of distance, the angular size of the half-light radius can be compared with the
size predicted by the FP to compute distance. As a distance estimator, the FP is accurate to within 15\%
\citep{saulder2013}. A more complete understanding of the evolution of \sig\ during mergers may allow the scatter in
the FP relation to be better understood.

The evolution of stellar velocity dispersion during mergers has only been previously studied in detail for a set of
highly idealized dissipationless merger simulations \citep[][hereafter denoted SC1]{stickley2012}. These simulations
suggested that \sig\ increases sharply whenever the nuclei of two progenitor galaxies pass through one another and
declines as the nuclei separate. As dynamical friction and tidal effects drive the nuclei toward coalescence, the time
between successive passes generally decreases. As a result, \sig\ undergoes damped oscillations of increasing frequency
preceding the final nuclear coalescence. After the nuclei coalesce, \sig\ undergoes much smaller, chaotic
oscillations as the system approaches a final state of equilibrium. However, the SC1 simulations did not include
gas dynamics, stellar formation, stellar evolution, rotating progenitors, disk galaxies, \smbhs, parent dark matter
halos, nor any feedback mechanisms.  Without including these effects, the results were not suitable for comparison to
real galaxy mergers. In the present work, we address the deficiencies of the SC1 simulations by performing a suite of
galaxy merger simulations that include all of these missing effects.

The research described in this paper was designed to aid in the interpretation of real galaxy mergers. When
possible, we have used analysis methods inspired by observational techniques and we have refrained from using certain
analysis techniques that are only possible or practical in numerical simulations. However, there is one major exception
to this rule; we use mass-weighted rather than flux-weighted measurements of \sig. In SC1, we found that the presence
of dust can, in principle, cause the flux-weighted value of \sig\ (i.e., the quantity measured in real galaxies) to
differ from its mass-weighted counterpart. We will characterize the difference between mass-weighted and flux-weighted
measurements of \sig\ in a subsequent paper.

The present paper is organized as follows. In Section 2, we describe the numerical simulations that we performed and
present details of the primary analysis routine. In Section 3, we present qualitative and quantitative results of the
simulations, including the temporal evolution of \sig\ and the evolution of \sig\ in various stellar populations. In
Section 4, we present additional statistical results regarding the intrinsic variability of \sig. We then
discuss the implications of our results and our planned future research in Section 5.

\section{NUMERICAL METHODS}

We performed a suite of binary galaxy merger simulations using the $N$-body, smoothed-particle hydrodynamics (SPH) tree
code, GADGET-3 \citep{springel2005}. Snapshots were saved at 5 Myr intervals, then each snapshot was analyzed
automatically using the analysis and visualization code, \GSnap\footnote{{http://www.gsnap.org}} (N.R. Stickley, in
preparation), which was designed for measuring velocity dispersions, computing statistics, and creating detailed volume
renderings of the gas and stars in $N$-body, SPH simulations of galaxies.

\subsection{The Simulation Code}

The stellar and dark matter particles in our simulations are simply treated as collisionless, gravitationally-softened
particles. The treatment of the gas component is considerably more complicated. \gadget\ simulates the hydrodynamics of
the interstellar medium (ISM) using a formulation of SPH that simultaneously conserves energy and entropy
\citep{springel2002}. The ISM is modeled as a multi-phase medium in which cold clouds are assumed to be embedded in a
hot, pressure-confining phase at pressure equilibrium \citep{springel2003}. The gas is able to cool radiatively and
become heated by supernovae. Consequently, the gas can convert between the hot and cold phases by condensing and
evaporating. Supernova explosions pressurize the ISM according to an effective equation of state parameterized by
$\qeos$ such that $\qeos=0$ corresponds to an isothermal gas with an effective temperature of $10^4$ K while $\qeos=1$
corresponds to the pure multi-phase model with an effective temperature of $10^5$ K. In the intermediate cases, $0 <
\qeos < 1$, the equation of state is a linear interpolation between the isothermal and multi-phase extremes.

\smbh\ feedback is modeled by treating each \smbh\ as a sink particle that accretes gas according to the
Bondi-Hoyle-Lyttleton parameterization,

\begin{equation}
 \dot M = \frac{4\pi\alpha G^2 M_{\rm BH}^2 \rho_\infty}{(c_\infty^2 + v_\infty^2)^{3/2}},\label{eq1}
\end{equation}

\noindent
where $\rho_\infty$ and $c_\infty$ are, respectively, the density and speed of sound in the local ISM and $v_\infty$ is
the speed of the \smbh\ relative to the local bulk motion of the ISM. The dimensionless parameter, $\alpha$, is a
correction factor introduced in order to account for the fact that the Bondi radius of the \smbh\ is smaller than the
resolution limit of the simulation. The bolometric luminosity of the accreting \smbh\ is $L = \epsilon_r \dot M c^2$,
where $\epsilon_r = 0.1$ is the radiative efficiency. A small fraction of the luminosity (5\% in our case) is assumed to
couple with the nearby surrounding gas (i.e., the gas within the \smbh's smoothing kernel), causing it to become heated.
The accretion rate is limited by the Eddington rate.

The star formation rate (SFR) depends on the density of the cool gas in the simulation. Specifically,
${\rm SFR}\propto\rho_{\rm sph}^{1.5}$, where $\rho_{\rm sph}$ is the density of the cool gas. The constant of
proportionality is chosen such that the simulated star formation rate surface density agrees with observations
\citep{kennicutt1998,cox2006}. In order to simulate basic stellar evolution, an instantaneous recycling approximation
is used; a fraction of the newly-formed stars is assumed to explode immediately as supernovae, enriching and heating
the surrounding ISM. Stellar wind feedback is simulated by stochastically applying velocity ``kicks'' to gas particles,
removing them from the dense star-forming region \citep{springel2003}. Mass is removed from the gas and used to create
new stellar particles. Each newly-created star particle carries with it a formation time variable. This makes it
possible to determine the age of each star particle that formed during the simulation.

\subsubsection{Simulation Parameters}

Our progenitor systems were constructed according to the method of \citet{springel2005a}. In summary, each system
contained a stellar bulge with a Hernquist density profile \citep{hernquist1990} of scale length, $R_{\rm bulge}$, and
an exponential disk of stars and gas. Each disk-bulge system was embedded in a dark matter halo with a Hernquist density
profile of scale length, $R_{\rm DM}$. A single \smbh\ particle was placed at the center of each system. In order to
test for stability, candidate progenitors were evolved forward in isolation; only stable systems were used in our merger
simulations. The details of each progenitor are presented in Table~\ref{table1}.

\begin{table*}
\begin{center}
\caption{\label{table1} Progenitor Galaxy Parameters}
\begin{footnotesize}
\begin{tabular}{cccccccccccc}
\tableline
\tableline
\vspace{-3mm}\\
Model                                   &
$N_{\rm part}$\tablenotemark{\it a}     &
$c$\tablenotemark{\it b}                &
\sig\tablenotemark                      &
$M_{\rm BH}$\tablenotemark{\it c}       &
$R_{\rm DM}$\tablenotemark{\it d}       &
$R_{\rm bulge}$\tablenotemark{\it d}    &
$R_{\rm disk}$\tablenotemark{\it e}     &
$M_{\rm total}$                         &
$M_{\rm bulge}$                         &
$M_{\rm disk}$                          &
$f_{\rm gas}$\tablenotemark{\it f}

\vspace{0.5mm} \\
                                        &
($10^{5}$)                              &
                                        &
($\rm km\, s^{-1}$)                     &
($\rm MM_\sun$)                         &
(kpc)                                   &
(kpc)                                   &
(kpc)                                   &
($\rm GM_\sun$)                         &
($\rm GM_\sun$)                         &
($\rm GM_\sun$)                         &
                                        \\
\tableline \vspace{-2.5mm}\\
0  & 16.08 [8.04+0+5.63+2.41]    & 9.5 & $89.9\pm1.7$ & 5.41 & 27.75 & 1.16 & 3.87 & 865.9 & 13.0 & 30.3 & 0.0 \\
A  & 16.08 [8.04+1.13+4.50+2.41] & 9.5 & $89.9\pm1.7$ & 5.41 & 27.75 & 1.16 & 3.87 & 865.9 & 13.0 & 30.3 & 0.2 \\
B  & 8.03 [4.02+0.61+2.44+0.96]  & 10.5& $68.5\pm2.1$ & 1.83 & 20.49 & 0.87 & 2.92 & 432.7 & 5.2  & 16.4 & 0.2 \\
C  & 4.02 [2.01+0.36+1.45+0.20]  & 12.5& $55.4\pm2.5$ & 0.51 & 14.30 & 0.64 & 2.12 & 216.2 & 1.1  & 9.7  & 0.2 \\
D  & 16.07 [8.04+0.56+5.06+2.41] & 9.5 & $89.9\pm1.7$ & 5.41 & 27.75 & 1.16 & 3.87 & 865.9 & 13.0 & 30.3 & 0.1 \\
E  & 16.08 [8.04+2.25+3.38+2.41] & 9.5 & $89.9\pm1.7$ & 5.41 & 27.75 & 1.16 & 3.87 & 865.9 & 13.0 & 30.3 & 0.4 \\
F  & 25.17 [12.58+1.88+7.50+3.21]& 9.5 & $89.9\pm1.7$ & 5.41 & 27.75 & 1.16 & 3.87 & 865.9 & 13.0 & 30.3 & 0.2 \\
\tableline
\end{tabular}
\end{footnotesize}
\end{center}
  \textit{a}The total number of particles [dark matter + gas + disk stars + bulge stars] in multiples of $10^5$.

  \textit{b} The concentration of the dark matter halo.

  \textit{c} The mass of the central black hole, measured in mega solar masses.

  \textit{d} Scale length of the Hernquist profile.

  \textit{e} Radial scale length of the stellar disk. The scale length of the gas disk is a factor
of six larger in each progenitor.

  \textit{f} The fraction of $M_{\rm disk}$ in the form of gas.

\end{table*}

We designed our suite of merger simulations to span a broad range of possible merger scenarios (see Table~\ref{table2}
for details). Our standard merger, labeled S1 in Table~\ref{table2}, was a tilted disk, prograde-prograde, 1:1 merger in
which the gas fraction in the disk of each progenitor was 0.2. In simulations S0--S7, we independently
varied the orbital parameters, mass ratios, and gas fractions in order to determine the effect of each property on the
evolution of \sig. Note that simulation~S0 contained no gas and was, therefore, a dissipationless merger. In
simulation S8, we varied the initial orbital parameters and increased the spatial and mass resolution of the
stars and gas particles by decreasing the gravitational softening length ($\epsilon$) and increasing the number of
particles, respectively. The initial masses of the \smbh\ particles were chosen to fall within the $1\sigma$ scatter of
the observed \msig\ relation from \citet{tremaine2002}.

In all simulations, the gravitational softening length of the dark matter and \smbh\ particles was 90 pc, the
accretion parameter, $\alpha$, from Equation~(\ref{eq1}) was set to 25, and we used $\qeos=0.25$. The simulations were
performed in a non-expanding space, rather than a fully cosmological setting.

\begin{table*}
\begin{center}
\caption{\label{table2} Merger Simulation Parameters}
\begin{footnotesize}
\begin{tabular}{cccccccccc}
\tableline
\tableline
\vspace{-2.5mm}\\
Simulation                          &
Progenitors                         &
Mass Ratio                          &
$r_0$\tablenotemark{\it a}          &
$r_{\rm min}$\tablenotemark{\it b}  &
$\theta_1$\tablenotemark{\it c}     &
$\phi_1$\tablenotemark{\it c}       &
$\theta_2$\tablenotemark{\it d}     &
$\phi_2$\tablenotemark{\it d}       &
$\epsilon$\tablenotemark{\it e}
\vspace{0.5mm}                      \\
                                    &
                                    &
                                    &
(kpc)                               &
(kpc)                               &
(deg)                               &
(deg)                               &
(deg)                               &
(deg)                               &
(pc)                               \\

\tableline \vspace{-3mm}\\
S0  & 0+0  & 1:1 & 150 & 5  & 25  & -20 & -25 & 20 & 25 \\
S1  & A+A  & 1:1 & 150 & 5  & 25  & -20 & -25 & 20 & 25 \\
S2  & A+A  & 1:1 & 150 & 5  & 205 & -20 & -25 & 20 & 25 \\
S3  & A+A  & 1:1 & 150 & 5  & 205 & -20 & 155 & 20 & 25 \\
S4  & B+A  & 1:2 & 150 & 5  & 25  & -20 & -25 & 20 & 25 \\
S5  & C+A  & 1:4 & 150 & 5  & 25  & -20 & -25 & 20 & 25 \\
S6  & D+D  & 1:1 & 150 & 5  & 25  & -20 & -25 & 20 & 25 \\
S7  & E+E  & 1:1 & 150 & 5  & 25  & -20 & -25 & 20 & 25 \\
S8  & F+F  & 1:1 & 120 & 10 & -30 &  0  & 30  & 60 & 20 \\
\tableline
\end{tabular}
\end{footnotesize}
\end{center}
\textit{a} The initial nuclear separation distance.

\textit{b} The nuclear pericentric distance of the initial orbit.

\textit{c} The initial orientation of galaxy 1. The angles $\theta$ and $\phi$ are spherical coordinates
measured in degrees, where $\theta=\arctan\left[(x^2+y^2)^{1/2}/z\right]$ is the inclination angle of the disk
with respect to the orbital plane, $\phi=\arctan(y/x)$, and the orbital plane is $z=0$.

\textit{d} The initial orientation of galaxy 2.

\textit{e} The gravitational softening length of stars and SPH particles. The softening length of dark matter
particles and \smbhs\ was 90~pc in all simulations.

\end{table*}

\subsubsection{Measuring \sig \label{analysis1}}

The primary quantity of interest, \sig, is the standard deviation
of the line-of-sight velocities of stars within the projected half-light radius of a galaxy's spheroidal component. In
practice, observational measurements of \sig\ are typically performed by placing a rectangular slit mask across the
center of the system in question to approximately isolate the half-light radius. The light passing through this slit
mask is then analyzed spectroscopically. In our analysis, \sig\ was measured using a method intended to mimic this
common observational technique. The \sig\ measurement algorithm, implemented within \GSnap, began by centering a virtual
rectangular slit mask of width $w$ and length $\ell$ on the galaxy of interest. A viewing direction, ($\theta, \phi$)
and slit position angle $\alpha$, were then chosen and the system was rotated such that the old ($\theta, \phi$)
direction corresponded with the new $z$-axis. The system was then rotated by $\alpha$ around the $z$-axis so that the
new $x$- and $y$-axes were parallel with the width and length of the slit, respectively. All stars appearing in the slit
were identified and stored in a list. Finally, the masses, $m_i$, and the line-of-sight component of the velocities,
$v_i$ of all stars in the list were used to compute the mass-weighted stellar velocity dispersion, \sig,

\begin{equation}
\sigmath = \sqrt{v_i^2m_i/M - (v_im_i/M)^2}
\label{eq2}
\end{equation}
\noindent
with
\begin{displaymath}
 M=\sum_i m_i
\end{displaymath}
\noindent
where the standard summation convention has been utilized; repeated indices imply a sum over that index.

No attempt was made to separate rotation from purely random motion. Consequently, measurements of \sig\ in a
dynamically cold rotating disk of stars yields larger values when measured along the plane of the disk than
when measured perpendicular to the disk. This choice was motivated by the fact that many observational measurements of
\sig\ are unable distinguish rotation from pure dispersion.

\subsubsection{Directional Statistics \label{analysis2}}

At 5 Myr intervals, \sig\ was computed along $10^3$ random directions, uniformly (i.e., isotropically) chosen
from the set of all possible viewing directions. For each viewing direction, a random slit mask position angle was
chosen uniformly from the interval $0\leq\alpha\leq\pi$ in order to simulate the effect of measuring \sig\ in randomly
oriented galaxies from random directions---just as is done when measuring \sig\ in real galaxies. Using this interval
potentially introduces a bias since slits oriented at $\alpha = 0$ and $\alpha = \pi$ are identical and are thus counted
twice. In practice, this bias was not detectable. Once the measurements of \sig\ were made, \GSnap\ computed the
directional mean, maximum, minimum, and standard deviation of \sig\ for the set of $10^3$ directions. When two
progenitor galaxies were present in the system, measurements of \sig\ were performed on only one of the progenitors. In
the two simulations containing progenitors of unequal mass, the measurements were centered upon the larger system.

\subsubsection{Precision}

Particle noise was the main source of uncertainty in our measurements of \sig. We quantified the uncertainty by first
constructing spherically symmetric particle distributions of the same size and mass as the galaxies that we were
analyzing. These particle systems were perfectly spherically symmetric---except for the statistical noise introduced by
using a finite number of particles, $N$. In the limit as $N\rightarrow\infty$, measurements of \sig\ in such systems
become independent of direction. Upon measuring the directional standard deviation of \sig\ (denoted \sigd) in these
spherical systems for various $N$ ranging from $10^3$ to $10^6$, we found the expected behavior: $\sigma_{\rm d}\propto
N^{-1/2}$. Determining the constant of proportionality associated with our simulation parameters allowed us to compute
the noise threshold associated with each individual measurement of \sig\ in each simulation. In our plots of \sig, the
uncertainty due to particle noise was comparable to the thickness of the plotted lines unless otherwise indicated in the
plot itself.

\subsubsection{Slit Size}

As mentioned previously, \sig\ is typically defined as the velocity dispersion of stars falling within the half-light
radius of the spheroidal component of a galaxy. In a disk galaxy containing a bulge, the starlight originating within
the half-light radius of the \textit{central bulge} is typically analyzed to obtain \sig. In elliptical systems, the
relevant light originates within the half-light radius of the \textit{entire system}. Of course, many systems do not
have well-defined spheroidal components. The lack of a spheroid makes it difficult to rigorously define
\sig---particularly in irregular galaxies---since measurements of \sig\ depend on the size of the slit. To simplify
matters, we have used a fixed slit of width $w = 2$~kpc and length $\ell=20$~kpc for all measurements of \sig\
throughout this paper. This rather large slit size, which corresponds to a slit of width $\approx 1''$ at a redshift of
0.1, allowed us to ensure that a large number of particles contributed to the measurement of \sig, thereby minimizing
particle noise. In our progenitor systems, this choice of slit size led to a systematic increase in the measured \sig\
of $\approx7\%$, compared with a slit that only included stars within the projected half-light radius of the bulge ($w =
0.3$~kpc, $\ell=3$~kpc). In our merger remnants, no difference was detected between slits measuring $2\times20$~kpc and
those measuring $0.3\times3$~kpc.

\section{RESULTS}

\begin{table*}
\begin{center}
\caption{\label{table3} Summary of Merger Characteristics}
\begin{footnotesize}
\begin{tabular}{cccccccccccc}
\tableline
\tableline
\vspace{-3mm}\\
Simulation                                  &
$t_1$\tablenotemark{\it a}                  &
$t_2$\tablenotemark{\it a}                  &
$t_3$\tablenotemark{\it a}                  &
$t_{\rm nc}$\tablenotemark{\it b}           &
$t_{\rm end}$\tablenotemark{\it c}          &
$\sigma_{\rm i}$\tablenotemark{\it d}       &
$\sigma_1$\tablenotemark{\it e}             &
$\sigma_2$\tablenotemark{\it e}             &
$\sigma_3$\tablenotemark{\it e}             &
$\sigma_{\rm final}$\tablenotemark{\it f,$\dagger$}&
\sigd\tablenotemark{\it g,$\dagger$}

\vspace{0.5mm}   \\

                                            &
(Gyr)                                       &
(Gyr)                                       &
(Gyr)                                       &
(Gyr)                                       &
(Gyr)                                       &
($\rm km\, s^{-1}$)                         &
($\rm km\, s^{-1}$)                         &
($\rm km\, s^{-1}$)                         &
($\rm km\, s^{-1}$)                         &
($\rm km\, s^{-1}$)                         &
($\rm km\, s^{-1}$)                        \\

\tableline \vspace{-2.5mm}\\
S0  & 0.37 & 2.09 & 2.33 & 2.50 & 3.83 & $89.9\pm10.1$ & 132.86  & 215.14 & 181.98 & $137.81\pm2.10$ & $11.37\pm0.61$\\
S1  & 0.37 & 2.06 & 2.30 & 2.40 & 3.62 & $89.9\pm10.1$ & 127.98  & 215.28 & 188.34 & $151.34\pm0.85$ & $6.59\pm0.30$\\
S2  & 0.37 & 2.17 & 2.40 & 2.47 & 4.09 & $90.2\pm10.3$ & 155.02  & 198.91 & 183.90 & $146.38\pm0.67$ & $5.04\pm0.23$\\
S3  & 0.37 & 2.09 & 2.31 & 2.45 & 3.40 & $89.6\pm9.7$  & 161.00  & 201.36 & 189.36 & $156.72\pm0.92$ & $5.52\pm0.19$\\
S4  & 0.42 & 2.04 & 2.30 & 2.40 & 3.73 & $89.6\pm9.3$  & 111.37  & 200.03 & 174.92 & $136.57\pm0.50$ & $8.19\pm0.26$\\
S5  & 0.46 & 2.37 & 2.71 & 2.89 & 4.69 & $89.9\pm9.2$  & 101.57  & 164.86 & 155.24 & $118.40\pm0.46$ & $7.94\pm0.31$\\
S6  & 0.37 & 2.09 & 2.32 & 2.40 & 3.59 & $89.3\pm10.3$ & 127.69  & 206.70 & 182.75 & $149.42\pm0.75$ & $6.74\pm0.29$\\
S7  & 0.37 & 2.03 & 2.26 & 2.35 & 3.54 & $90.9\pm8.5$  & 129.66  & 206.76 & 187.24 & $146.85\pm0.98$ & $6.84\pm0.61$\\
S8  & 0.27 & 3.28 & 3.57 & 3.72 & 4.79 & $86.3\pm5.1$  & 95.5    & 224.37 & 189.71 & $152.42\pm0.37$ & $5.57\pm0.25$\\
\tableline
\end{tabular}
\end{footnotesize}
\end{center}

\textit{a} The time of the \textit{n}th pass.

\textit{b} The time of nuclear coalescence.

\textit{c} The duration of the simulation. This indicates the time at which the snapshots in Figure
\ref{fig3} were saved. This is the only quantity in the table that does not depend upon the initial conditions of each
merger.

\textit{d} The velocity dispersion of the progenitor system. The reported uncertainty is the standard
deviation of the set of $10^3$ random measurements.

\textit{e} The mean value of \sig\ at the climax of the \textit{n}th pass (i.e., the value at time
$t_n$).

\textit{f} The mean velocity dispersion of the remnant system.

\textit{g} The standard deviation of the directional distribution of \sig.

\textit{$\dagger$} This was obtained by averaging the time series over the final 500 Myr of the simulation. The
reported uncertainty is the standard deviation of the time series during the 500 Myr interval.
\end{table*}

\subsection{Merger Evolution}

\begin{figure*}
    \includegraphics[width=6.9in]{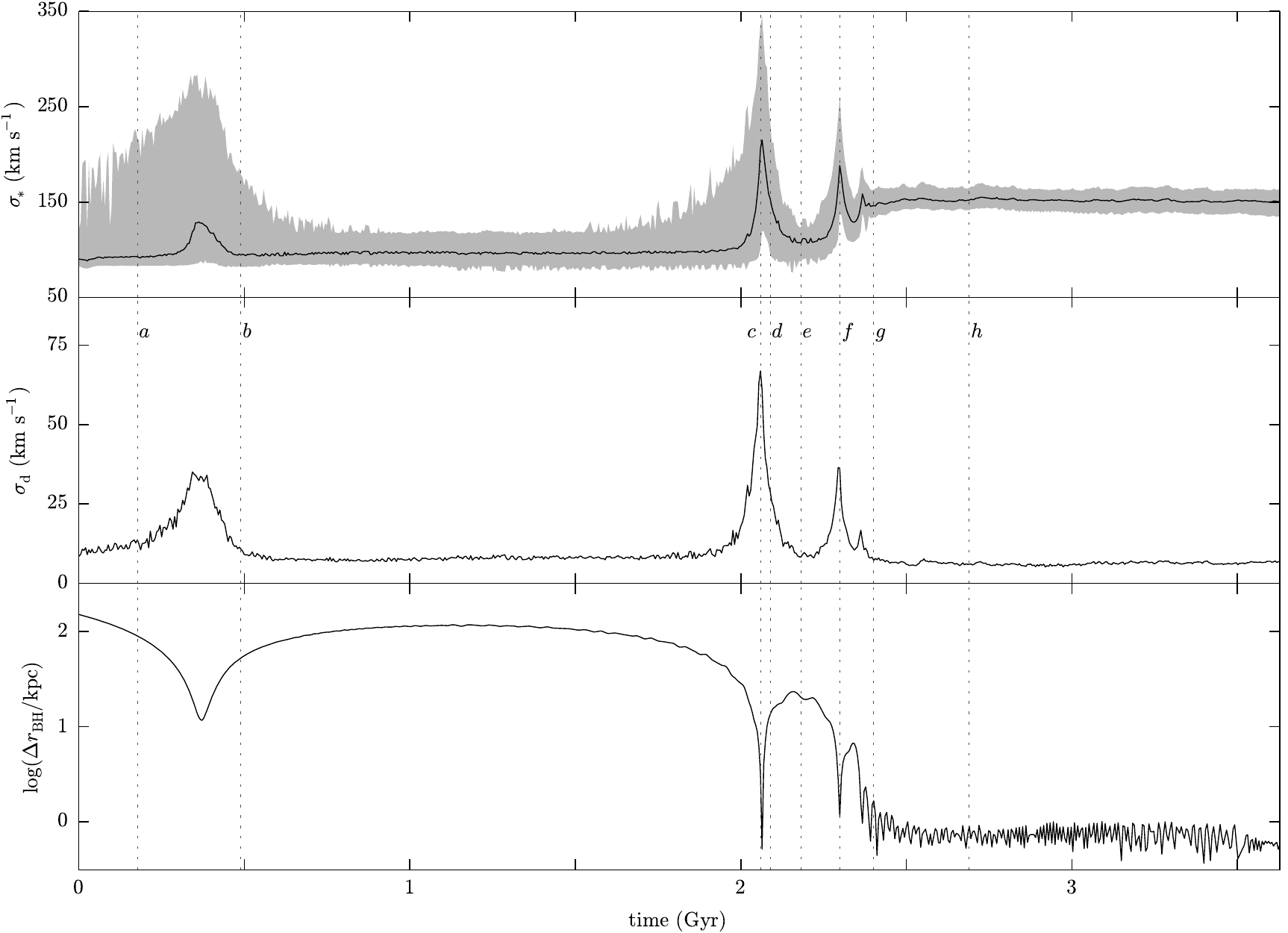}
    \caption{\label{fig1}\footnotesize Merger evolution time series for simulation S1.
    Upper panel: The mean value of \sig\ over the set of 1000 viewing directions is plotted in black. The upper
    and lower edges of the gray shaded region show the maximum and minimum values of \sig.
    Middle panel: The standard deviation of the set of \sig\ measurements.
    Lower panel: The distance between the two \smbh\ particles. This is a proxy for the distance between the nuclei of
    the two progenitors. The dotted vertical lines indicate the time coordinates of snapshots that are examined
    in further detail later in this paper. Visual renderings of these snapshots are presented in Figure~\ref{fig2}.
    The snapshots are located at $a = 0.177$ Gyr, $b = 0.490$ Gyr, $c = 2.060$ Gyr, $d = 2.089$ Gyr, $e = 2.181$
    Gyr, $f = 2.299$ Gyr, $g = 2.401$ Gyr, and $h = 2.690$ Gyr.}
    \vspace{2mm}
\end{figure*}

\begin{figure*}

    \vspace{-0.66in} 

    \includegraphics[width=6.9in]{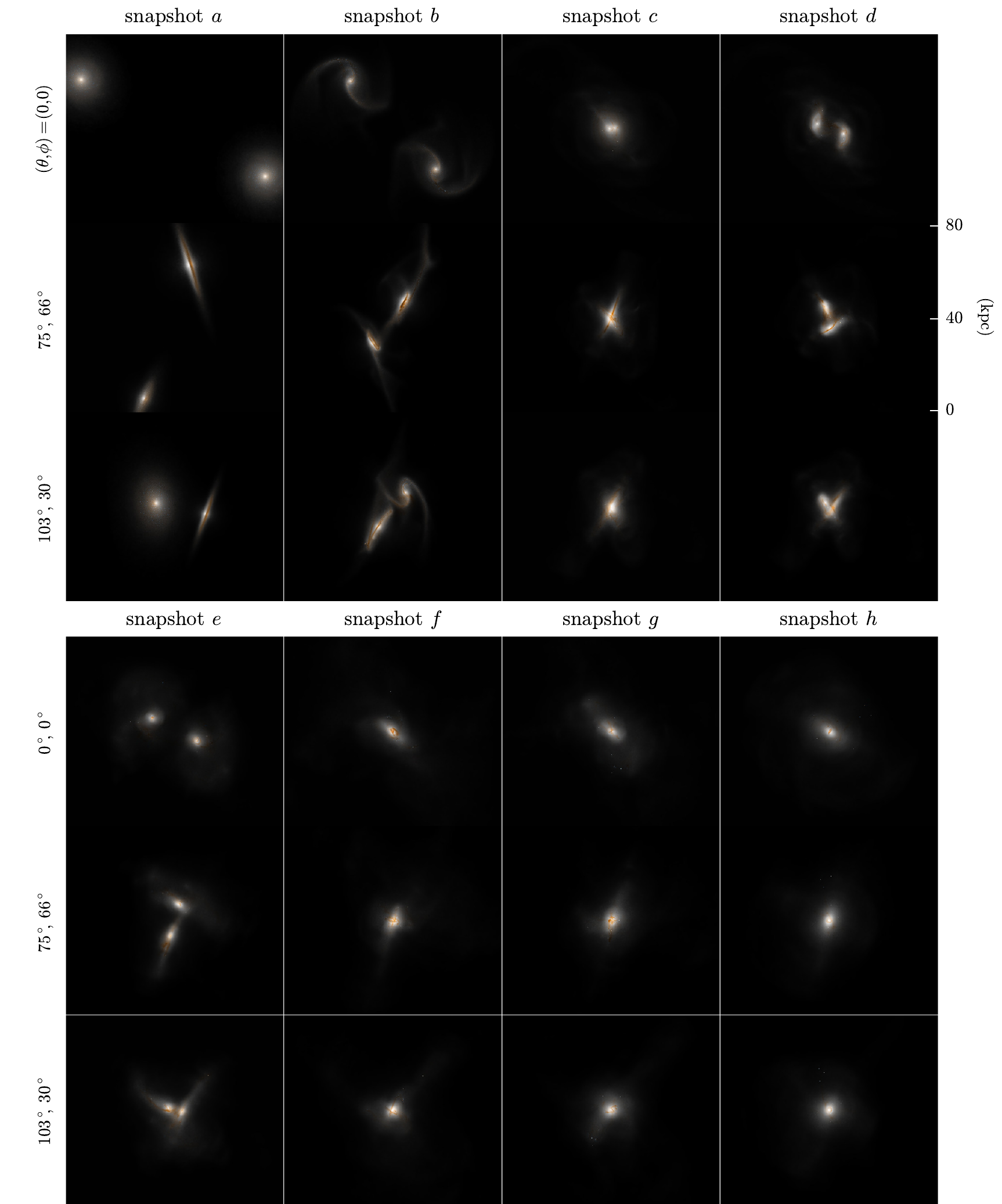}
    \caption{\label{fig2}\footnotesize Visualizations of snapshots from simulation S1, created using \GSnap's volume
    rendering algorithm. The snapshot times correspond to the dotted vertical lines in Figure~\ref{fig1}. Each
    snapshot is shown from three directions, indicated in spherical coordinates on the left. The width and height of
    each image is 93.75 kpc and 81.19 kpc, respectively.}
\end{figure*}

In order to better understand the following discussion, it will be helpful to refer to Figures~\ref{fig1} and
\ref{fig2}. In Figure~\ref{fig1}, we present time series data for \sig\ and the \smbh\ separation distance during the
1:1 merger, S1. Vertical lines indicate key moments in the evolution of the system. Images of the system during these
moments, or ``snapshots,'' are shown in Figure~\ref{fig2}. Note that whenever the word ``nucleus'' is used in
this paper, we are referring to the position of one of the local maxima in the density field of the system. Nuclei also
coincide with local minima in the smoothed gravitational potential field, but nuclei do not necessarily coincide with
the positions of \smbhs. When we say that nuclei have coalesced, we mean that two local minima in the gravitational
potential field have combined to form a new, deeper global minimum that persists indefinitely.

As described above, each dissipative simulation (S1--S8) began with two disk galaxy progenitors composed of a
central bulge, a stellar disk, a thin disk of gas, a dark matter halo, and a central \smbh. The exact details of each
merger, listed in Table \ref{table1}, varied, but they all of the dissipative mergers shared the following qualitative
features: As soon as the simulations began, star formation commenced. A spiral density pattern developed in the gas
component of each progenitor. Enhanced star formation in the dense regions of gas led to a spiral pattern in the
distribution of new stars. As the density of the spiral arms increased, the preexisting population of older disk stars
gradually began participating in the spiral pattern, but only slightly. Lacking gas, the dissipationless
simulation was unable to form stars. No spiral density pattern developed in the dissipationless disk progenitors.

The parent dark matter halos of the progenitors initially overlapped somewhat, however the stellar components were
initially significantly separated. The progenitors followed approximately parabolic orbits while approaching one
another. Tidal forces grew stronger and began to visibly elongate and warp the progenitors as they prepared to collide.
Shortly before reaching their pericentric distance, the galaxies began to overlap significantly and \sig\ began
increasing. Simultaneously, the standard deviation of the \sig\ distribution (\sigd) increased. The gas components of
the progenitors collided and became compressed. A small fraction of the
gas lost enough angular momentum in this initial collision to begin migrating toward the nuclei of the progenitors. As
the nuclei reached the pericentric distance, \sig\ reached a maximum value. This increase in \sig\ was
primarily due to
the projected streaming motion of the progenitors rather than a true increase in \sig; lines of sight perpendicular to
the collision axis experienced very little enhancement in \sig, while lines of sight coinciding with the collision axis
(i.e., lines of sight along which stars of both bulges simultaneously fell within the measuring slit) yielded the
largest values of \sig.

While receding from the first encounter, the velocity dispersion of each progenitor quickly returned to its
pre-collision value. Strong tidal tails and a bridge of stars and gas began forming. A small amount of gas finally
reached the nuclei and triggered short, sporadic episodes of AGN activity upon reaching the \smbhs. Gas in the tidal
tails collapsed to form thin filaments as the galaxies continued to recede. The filaments then
fragmented to form clumps from which clusters of stars soon formed as discussed in \citet{elemgreen1993},
\citet{barnes1996}, \citet{wetzstein2007}, and references therein. The approximate time of the fragmentation and
cluster formation in merger S1 is marked by snapshot \textit{b} in Figure~\ref{fig1}. These clusters, which can be seen
in Figure~\ref{fig2}
and Figure~\ref{fig3} as bright compact spots, most likely represent tidal dwarf galaxies. With diameters of 50--300 pc
and masses of $10^7$--$10^9\,\rm M_\sun$, these systems lie near the resolution limit of our simulations; some of the
smaller ones may merely indicate the formation sites of small structures such as globular clusters. Observational
evidence for such tidal dwarf galaxies is reviewed in \citet{dabringhausen2013}. Although no tidal dwarf systems
formed in our dissipationless merger, we note that it is possible for tidal dwarf systems to form in dissipationless
mergers at this stage \citep[see][for details]{barnes1992}.

After receding from one another, the progenitors eventually reversed direction and began approaching one another on a
trajectory that was much more nearly head-on than the first approach. Upon the second approach, the interstellar gas of
the two progenitors collided once again (snapshot \textit{c}), losing considerably more angular momentum this time. In
contrast with the first encounter, \sig\ increased along all lines of sight; the minimum, maximum, mean, and standard
deviation of \sig\ increased sharply as the nuclei passed through one another and then decreased as the nuclei receded
(see snapshots \textit{d} and \textit{e}). As the nuclei reversed direction again, \sig\ nearly returned to its initial
value. Simultaneously, in-falling clumps of low-angular momentum gas began reaching the central \smbhs, triggering
significant episodes of AGN activity. The nuclei then began approaching one another while stars in the outer regions of
the merging system, where the dynamical timescale was longer and the stars were less tightly bound, continued on nearly
the same trajectories that they followed during the second approach---essentially unaffected by the motion of the
nuclei.

As the nuclei began to overlap for the third time, AGN activity decreased significantly and \sig\ increased once again
(snapshot \textit{f}). After passing through one another once more, the velocity dispersion of each nucleus decreased
somewhat, but it did not return to its initial value. This process was repeated several more times in rapid succession,
with more stars being shed from the nuclei during each reversal. The nuclear turnaround distance decayed until the two
nuclei eventually coalesced (at snapshot \textit{g}). Oscillations in the value of \sig\ decayed away during this stage
and the system adopted a new, stable \sig. During the final stages of nuclear coalescence, gas and stars of low angular
momentum began falling into the deep potential well of the new nucleus. This triggered a nuclear starburst which was
soon followed by the highest \smbh\ accretion rates of the entire merger process. The accretion episodes during this
stage were more frequent and more sustained than at any other time during the simulations (see images of
snapshot~\textit{g} for the corresponding morphology). The surrounding gas became heated by the AGN, expanded, and
drove significant gas outflows \citep[see][for a detailed discussion of these phenomena]{hopkins2006}. The stars that
fell toward the nucleus soon passed through the nucleus and emerged in spherical waves on the other side only to fall
back onto the nucleus again. The effect of the stars falling toward the nucleus, overshooting, and then falling back
caused small, statistically significant fluctuations in \sig---the same oscillations that were observed in the ``phase
mixing'' merger stage described in SC1. The amplitude of these oscillations gradually decreased as the system became
more throughly mixed. Stars that were ejected after the second and third passes gradually fell back toward the nucleus
during the $\approx 1$~Gyr following the final coalescence. In all of our dissipative simulations, clumps of
gas that were ejected without being significantly heated earlier in the merger process also fell back toward the nucleus
and formed a series
of nuclear disks with diameters ranging from 100~pc to 10~kpc (disks smaller than 100~pc could not be resolved). The
formation of similar disks is discussed in \cite{barnes2002}. Gas of sufficiently low angular momentum was able to
accrete onto the \smbh(s), causing another period of significant AGN activity $\gtrsim 1$~Gyr after final coalescence.
This late-stage accretion was observed in all of our mergers except for the lowest gas mass fraction mergers, S0 and S6.
In mergers that still contained two distinct \smbhs\ at this late stage, the formation of the nuclear disks allowed for
efficient angular momentum transfer from the \smbhs\ to the disk material, as discussed in \citet{gould2000} and
\citet{escala2005}. In real galaxies, the \smbhs\ have most likely merged before this late stage; the spatial resolution
of our simulations was insufficient to follow the details of the binary \smbh\ orbital decay \citep{escala2005}, thus
the \smbh\ merger timescale could not be accurately simulated. Images of the final remnant galaxies are presented in
Figure~\ref{fig3}.

\subsection{Dependence upon Initial Parameters}

The general shapes of the three time series shown in Figure~\ref{fig1} are shared by all of our
mergers---including the dissipationless merger, S0. Rather than presenting plots for each merger, we have summarized
the basic features of each merger in Table \ref{table3}. We report the time coordinates of the first three passes
($t_1$--$t_3$), the mean stellar velocity dispersion of the systems during each pass ($\sigma_1$--$\sigma_3$), the time
at which the nuclei coalesced ($t_{\rm nc}$), the value of \sig\ in the remnant, and the duration of each simulation.

Simulations S1, S2, and S3 tested the dependence of \sig\ upon the spin-orbit configuration of the initial system. From
the data, it appears that configurations of lower net angular momentum cause more significant increases in \sig\ during
the first encounter; the high angular momentum, prograde-prograde merger (S1) exhibited the lowest $\sigma_1$ value,
while the merger of lowest angular momentum (S3) exhibited the highest value of $\sigma_1$. No other trends were
observed with respect to the spin-orbit configuration.

\begin{figure*}
\vspace{-0.7in} 

    \includegraphics[width=6.9in]{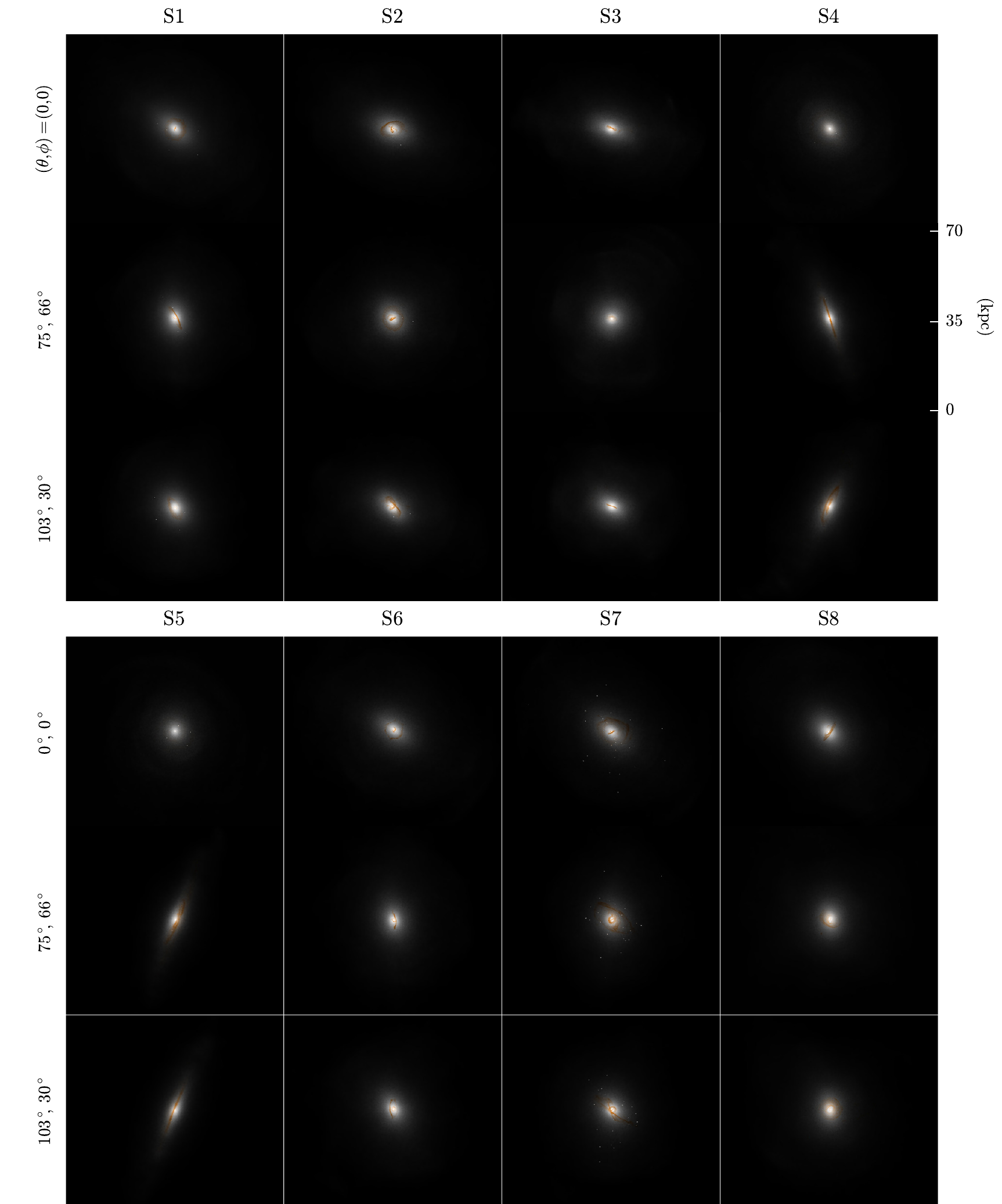}
    \caption{\label{fig3}\footnotesize The remnants of merger simulations S1--S8, listed in Table \ref{table2}. The
    simulation time of each snapshot is listed in Table~\ref{table3}, as $t_{\rm end}$. Each remnant is viewed from
    the same three directions as the images in Figure~\ref{fig2}. Each field of view measures 84.38~kpc~$\times$~
    73.071 kpc. All remnants are elliptical except for S4 and S5---the unequal mass mergers. All systems contain
    nuclear disks.}
\end{figure*}

The gas fraction of the disk component was varied from 0.0 to 0.4 in simulations S0, S1, S6, and S7. The
elapsed time between the first and second encounters was mildly dependent on the gas fraction, with higher gas fractions
leading to shorter intervals. This was likely caused by the dissipative, collisional nature of the gas; when more gas
was present, translational kinetic energy was more efficiently converted into internal energy, resulting in slightly
lower recession velocities. The star formation rate was higher in mergers with larger gas fractions, since these systems
contained more raw material from which to build stars. In Figure~\ref{fig3}, we see that the number of tidal dwarf
galaxies also increased with gas fraction. The presence of more dwarfs made the time series data slightly
more noisy as the gas fraction increased from 0.1 to 0.4. However, note that the remnant of the completely
dissipationless merger (i.e., the gas-free merger, S0) exhibited the least stable value of $\sigma_{\rm final}$, even
though no tidal dwarf systems formed. This was apparently due to the phase mixing process, described in SC1, which was
more pronounced in the absence of dissipation. We found no clear relationship between gas fraction and $\sigma_{\rm
final}$.

In simulations S1, S4, and S5, the mass ratio of the progenitors was varied. Unsurprisingly, systems of comparable mass
were able to disturb one another more effectively. This led to more significant enhancements in \sig\ during the merger
process. The other trends evident in S1, S4, and S5 can be attributed to the varying total masses of these systems; the
systems of higher total mass merged more rapidly and produced systems of higher \sig.

In simulation S8, as well as many low-resolution trial simulations, the orbital parameters were varied. Smaller
pericentric distances lead to faster mergers and larger enhancements in \sig\ during the first pass. In the case of
nearly head-on initial encounters, \sig\ reaches its absolute highest value during the first pass rather than the
second pass.

\subsection{The Distribution of \sig\label{sect_dist}}

\begin{figure*}

    \includegraphics[width=6.9in]{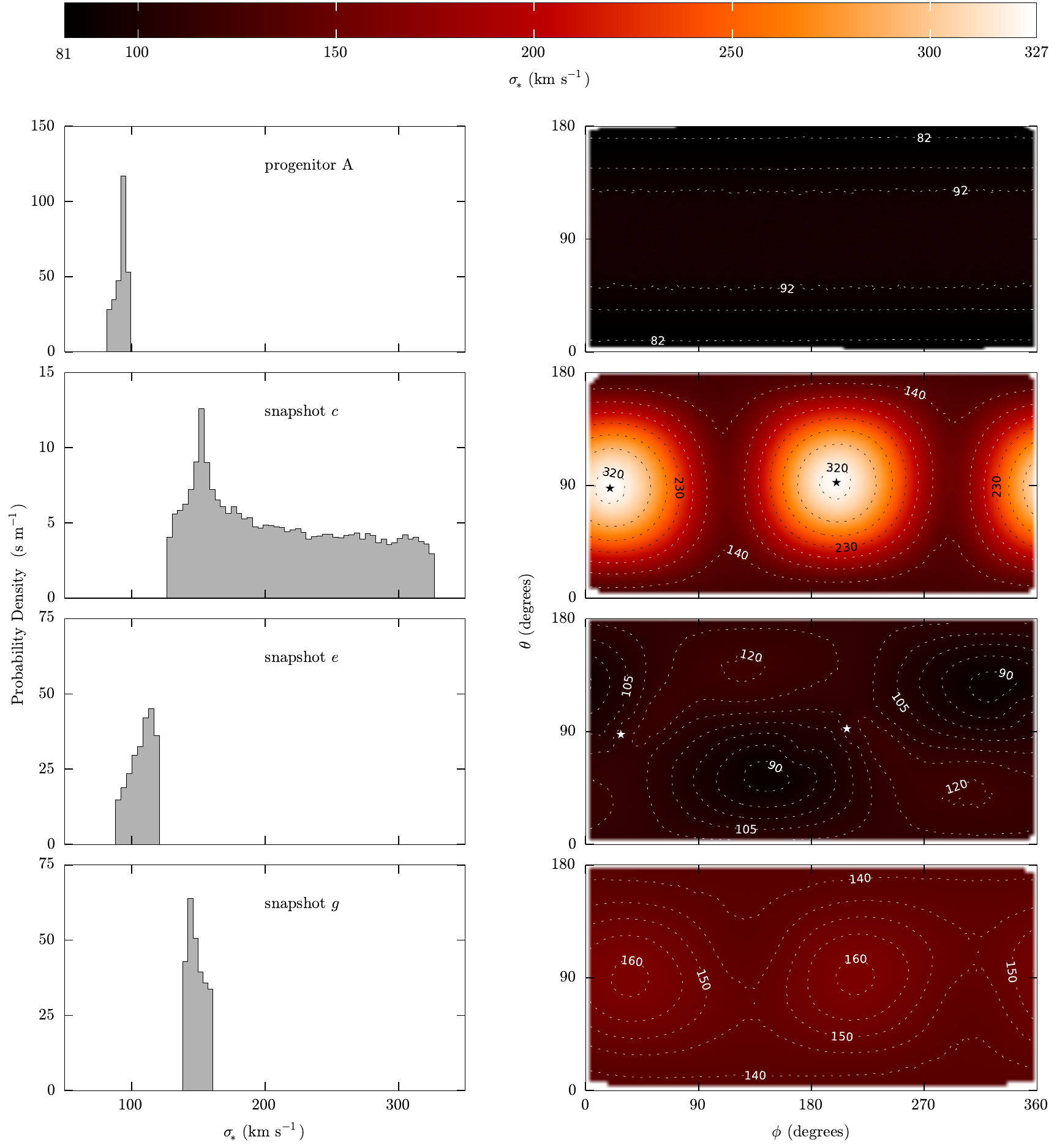}

    \caption{\label{fig4}\footnotesize The angular and probability distributions of \sig\ in progenitor galaxy
    A (upper panel), followed by snapshots \textit{c}, \textit{e}, and \textit{g} of simulation S1 (see Figure
    \ref{fig1} for more information on the meaning of these labels). For each snapshot, \sig\ was measured along 20,000
    random directions. The contour plots on the right show the directional variation of \sig, while the histograms on
    the left show the corresponding probability distributions. The star symbols indicate the directions that lie along
    the instantaneous collision axis, where applicable. For a discussion of this figure, see section \ref{sect_dist}.}
\end{figure*}

While the time series presented in Figure~\ref{fig1} are helpful for understanding the evolution of \sig\ with time,
they do not contain much information regarding the distribution of \sig\ during the merger process. To supplement the
time series data, we present, in Figure~\ref{fig4}, the angular and probability distributions of \sig\ in four
snapshots during merger S1. For each of these four snapshots, \sig\ was measured along 20,000 lines of sight.

In the progenitor galaxy, \sig\ is distributed nearly isotropically. The presence of a stellar disk is
evident from the symmetry about the equator of the system ($\theta=90^\circ$). Measurements of \sig\ along lines of
sight perpendicular to the disk are diminished by the presence of the disk stars while measurements made along the
edge of the disk are enhanced somewhat because, from these sight lines, the disk's rotation can contribute to the
velocity dispersion measurements.

The second snapshot of interest (snapshot~\textit{c}) was recorded shortly before the climax of the second encounter. It
shows that the highest values of \sig\ are measured along the collision
axis (denoted by the star symbols) while the lowest values are measured perpendicular to the
axis. Furthermore, the positive skew of the probability distribution indicates that a random measurement of \sig\ during
a collision is more likely to yield a value near the mean or minimum rather than near the maximum possible value
($\sigma_{\!*,\rm max}$). The reason \sig\ is highest along the collision axis is twofold: (1) the two progenitors are
moving with respect to each other along this axis; their combined bulk motion is mistaken for stellar velocity
dispersion when the system is viewed along this direction, and (2) the actual velocity dispersion of each progenitor
increases along the collision axis during collisions. These separate effects are discussed in more detail in the next
section.

Midway between the second and third passes (snapshot~\textit{e}), the lines of sight yielding the maximum measurements
of \sig\ no longer coincide with the instantaneous collision axis. We can also see that a random measurement of \sig\ is
more likely to fall near the maximum value than near the minimum. However, the system is considerably more isotropic
than it was during the climax of the second pass, so the difference between the minimum and maximum values of \sig\ is
much less significant here.

Immediately after nuclear coalescence (snapshot~\textit{g}), \sig\ is already distributed quite uniformly; the
difference between the maximum and minimum \sig\ is much smaller than during snapshot~\textit{c}. The contours in the
angular distribution plot for snapshot~\textit{g} indicate that the velocity dispersion is highest along a preferred
axis---similar to the distribution in snapshot~\textit{c}. This axis corresponds to the collision axis during the last
few encounters before coalescence, which is not necessarily the same as the collision axis during the second pass.

\subsection{Random versus Streaming Motion\label{sect_random}}

The measurements of \sig, discussed above, have been based upon a straightforward application of Eq. (\ref{eq2}) to all
stars appearing in a slit mask centered on one of the nuclei of a merging system. Since this is the
observationally accessible quantity, it would be more appropriate to refer to this version of \sig\ as the
\textit{apparent} velocity dispersion. The apparent velocity dispersion includes the effects of rotation and bulk
motion whereas the \textit{intrinsic} velocity dispersion is due to the purely random motion of stars in the system.

Suppose two systems with intrinsic velocity dispersions $\sigma_1$ and $\sigma_2$ move toward or away from one another
with speed $v$. Let $m_1$ and $m_2$ be the portions of the stellar masses of systems 1 and 2 that appear within a slit.
Using Eq. (\ref{eq2}), it is possible to show that the apparent velocity dispersion along the line of
sight connecting the centers of two systems is given by,

\begin{equation}
\sigmath = \sqrt{f_1\sigma_1^2 + f_2\sigma_2^2 + f_1f_2v^2 },
\label{eq3}
\end{equation}

\noindent
where $f_i$ are the fractional masses,
\begin{displaymath}
 f_i = \frac{m_i}{m_1+m_2}.
\end{displaymath}

\noindent
For the special case of two identical systems of velocity dispersion $\sigma_0$ on a collision course, a measurement
of \sig\ along the collision axis will yield

\begin{equation}
\sigmath = \sqrt{\sigma_0^2 +(v/2)^2 },
\label{eq4}
\end{equation}
\noindent
since $m_1=m_2$ and $\sigma_1=\sigma_2=\sigma_0$.

In major mergers, the \sig\ appearing on the left side of Equations (\ref{eq3}) and (\ref{eq4}) typically corresponds
to the maximum measurement of velocity dispersion in the merging system. Thus, $\sigma_{\!*,\rm max}$ can be used as an
approximation for this quantity. The relative radial speed of the two systems, $v$, can be approximated using the
relative radial speed of the two \smbhs. More precisely, if $\mathbf{r}$ is a position vector pointing from one black
hole to the other and $\mathbf{v}$ is the corresponding relative velocity vector, the speed $v$ is given by
$v=|\hat\mathbf{r}\cdot\mathbf{v}|$, where $\hat\mathbf{r} =\mathbf{r}/|\mathbf{r}|$.

\begin{figure}
    \includegraphics[width=3.3in]{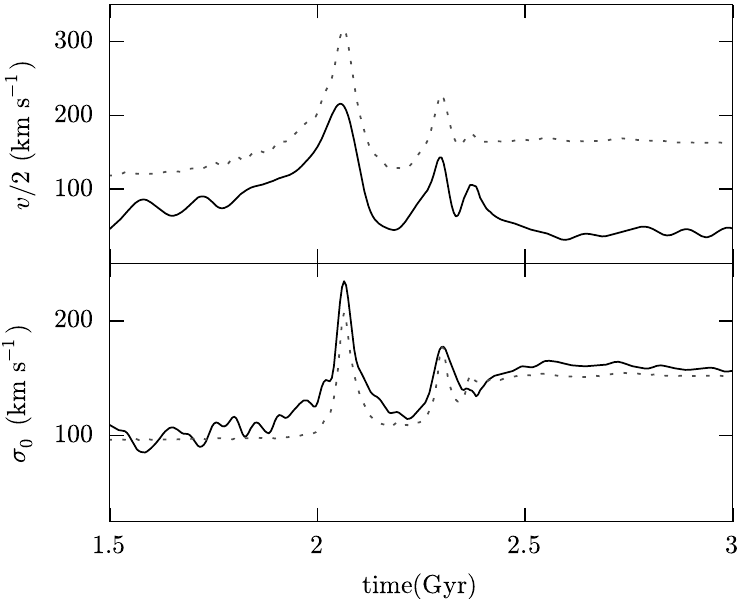}
    \caption{\label{fig5}\footnotesize Separating the intrinsic and apparent velocity dispersions. Upper panel: The
    solid black line shows $v/2$, defined in Section \ref{sect_random}. The gray
    dashed line shows the maximum apparent velocity dispersion ($\sigma_{\!*{\rm,max}}$) as a function of time during
    merger S1. Both quantities have been smoothed over time to remove high frequency fluctuations.
    Lower panel: The solid line shows the intrinsic velocity dispersion, $\sigo$, discussed in Section
    \ref{sect_random}. The dashed line shows the mean value of \sig\, over the set of 1000 viewing directions;
    it is a smoothed version of the plot in the upper panel of Figure~\ref{fig1}. Note that the mean velocity
    dispersion closely traces the intrinsic velocity dispersion.}
\end{figure}

In the special case of a merger of identical systems, measuring $\sigma_{\!*,\rm max}$ and $v$ allows us to infer the
intrinsic velocity dispersion ($\sigo$) of each system, using Equation~(\ref{eq4}). In Figure~\ref{fig5}, we show the
result of decomposing measurements of apparent velocity dispersion into streaming and intrinsic components. The analysis
was performed on merger S1, which began as a merger of identical systems. The upper panel shows $\sigma_{\!*,\rm max}$
and $v/2$ while the lower panel shows the intrinsic velocity dispersion, $\sigo$, measured along the collision axis.
The results suggest that the mean of \sig\ over all directions (the dashed line in the lower panel) closely
traces the intrinsic velocity dispersion (the solid line). The intrinsic velocity dispersion along the collision axis
is only mildly elevated in comparison with the mean value of \sig. However, there are several caveats: First, we note
that the velocity of a \smbh\ does not trace the velocity of its parent nucleus perfectly. In general, each \smbh\
particle orbits the center of its parent nucleus. We have smoothed the $v$ time series in order to remove high frequency
variations caused by this motion. For consistency, we also smoothed the $\sigma_{\!*,\rm max}$ time series. Secondly,
the velocity of a nucleus does not always trace the bulk velocity of its parent progenitor galaxy. In fact, neither
progenitor galaxy has a well-defined bulk velocity during a collision; as the progenitor systems become increasingly
superimposed, the streaming velocity in each progenitor begins to vary with position. Finally, even though
$\sigma_{\!*,\rm max}$ is usually a good approximation for the quantity on the left side of Equation~(\ref{eq4}), this
is not necessarily true at the turnaround times when the streaming velocity is low. In such cases, the maximum velocity
dispersion is not necessarily measured along the collision axis (see Figure~\ref{fig4}). In light of these complexities,
it would be best to interpret the resulting plot of $\sigo$ qualitatively rather than quantitatively.

\subsection{Evolution with Stellar Age}

It has long been known that the velocity dispersion of stars in the disk of the Milky Way increases with age. This
so-called ``age-velocity relation,'' along with the phenomena which cause it, have been studied for more than six
decades \citep{spitzer1951,spitzer1953,barbanis1967,hanninen2002,nordstrom2004}. More recently, it has been shown that
measurements of \sig\ in more distant galaxies can depend upon the population of stars being measured
\citep{rothberg2010,rothberg2013}. Specifically, measurements of \sig\ that are based upon the spectral features of
younger stars yield lower values than measurements of \sig\ which include all stars or only older K and M stars. To
explain this ``$\sigma$-discrepancy,'' Rothberg et al. argue that stars are born with low velocity dispersion, since the
gas from which stars form is dynamically cold, due to its
dissipative, collisional nature.

In order to investigate whether our simulations exhibited age-dependent \sig, we performed the analysis described in
sections \ref{analysis1} and \ref{analysis2} using only stars in specified age ranges. Before
presenting our findings, though, we note that the results presented in this section necessarily depend strongly upon
the less robust aspects of the simulation code---namely, the numerical methods used to simulate the hydrodynamics, star
formation, and \smbh\ feedback. These methods effect the timing, location, and rate of the star formation. Furthermore,
the particles traced by our simulations do not represent individual stars. Instead, they represent small regions of star
formation. This means that cluster evaporation and other small-scale effects are not included. Therefore, our results
regarding the evolution of \sig\ with stellar age are less robust than our age-independent analysis.

\begin{figure}
    \includegraphics[width=3.3in]{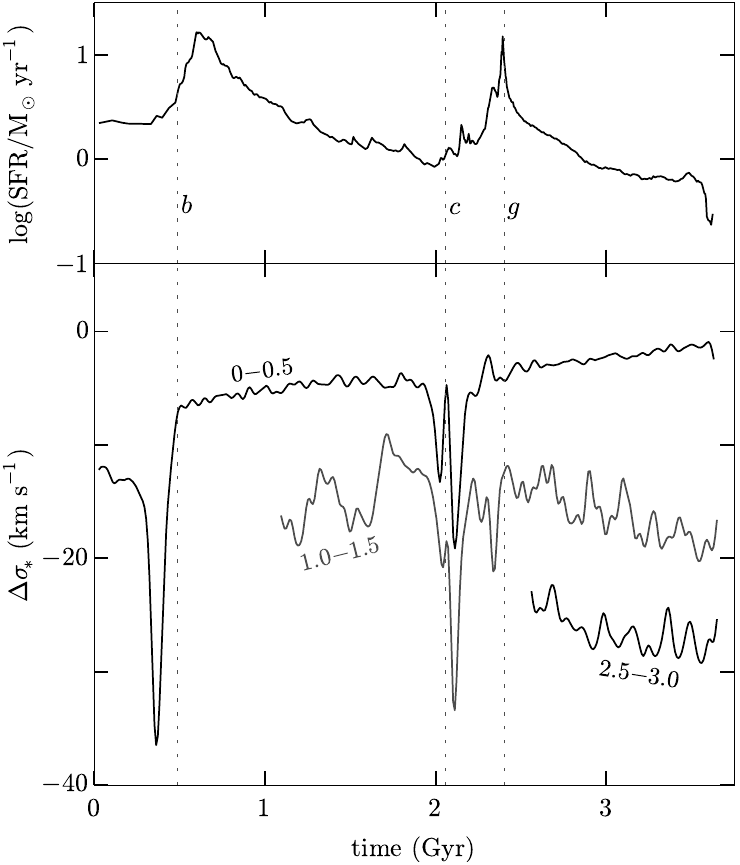}
    \caption{\label{fig6}\footnotesize The evolution of \sig\ for various stellar age bins. Upper panel: The star
    formation rate in simulation S1. Lower panel: The offset from the global, mean velocity dispersion that includes
    stars of all ages, $\Delta\sigmath = \sigmath - \sigma_{\!*,\rm all}$ for three stellar age bins of width 0.5
    Gyr. Each line is labeled with its age bin. For example, the top line shows the evolution of $\Delta\sigmath$
    for all stars that were born between time $t=0$ Gyr and $t=0.5$ Gyr. The vertical dashed lines indicate snapshot
    times introduced in Figure \ref{fig1}.}
\end{figure}

In Figure~\ref{fig6}, we present the evolution of \sig\ for stars in three age bins during simulation S1 by plotting the
offset from the instantaneous global value of \sig\ (i.e., the value of \sig\ based on stars of all ages). The star
formation rate is plotted in the same figure for reference.

Stars that formed during the first 0.5~Gyr of the simulation
were located in the disks of the progenitor systems. These stars were born with $\sigmath\approx 12 \rm\,km\,s^{-1}$
lower than the global velocity dispersions of their parent galaxies. Immediately after the first pass, the offset was a
mere $\approx 7\,\rm km\,s^{-1}$. These stars gradually mixed and became dynamically heated. By the end of the
simulation, they were essentially dynamically indistinguishable from the system as a whole.

The evolution of stars that formed between 1.0~Gyr and 1.5~Gyr (i.e., between the first and second passes) after the
beginning of the simulation was more complicated because approximately 70\% of these stars formed in tightly-bound tidal
tidal dwarf-like systems. The dwarf galaxies repeatedly passed near the nuclei of the larger systems. This lead to large
fluctuations in the \sig\ evolution time series. While the two primary galaxies were approaching one another, in
preparation for their second encounter, \sig\ in this age bin generally increased. However, after the second pass, \sig\
began decreasing; the offset from the global \sig\ increased while the global value remained essentially constant, as
seen in Figure~\ref{fig1}. This behavior is due to the orbital decay of the satellite galaxies in which these stars are
primarily located.

Finally, stars that were born between 2.5~Gyr and 3.0~Gyr (i.e., immediately following nuclear coalescence) formed
exclusively in the nuclear disks and satellite galaxies (in the case of simulation S1, 85\% formed in the nuclear disks
while the remaining 15\% formed in the satellite galaxies). The behavior of \sig\ in this group was similar to the
1.0~Gyr to 1.5~Gyr group, although the offset was larger by $\approx 10 \rm\,km\,s^{-1}$.

\begin{figure}
    \includegraphics[width=3.3in]{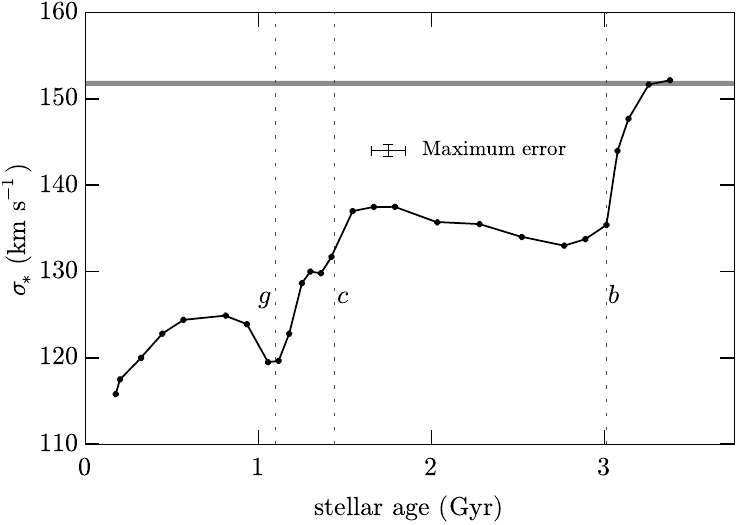}
    \caption{\label{fig7}\footnotesize The relation between \sig\ and stellar age in simulation S1. The snapshot
    examined here was saved at 3.5 Gyr (i.e., stars that formed at $t=0$~Gyr are 3.5~Gyr old). The vertical dashed lines
    indicate the ages of stars that were born at the corresponding snapshot times shown in Figures \ref{fig1} and
    \ref{fig6}. The horizontal bar indicates the global value of \sig\ in the current snapshot.}
\end{figure}

In Figure \ref{fig7}, we have plotted \sig\ as a function of stellar age in the remnant system in a simulation snapshot
that was saved at $t=3.5$~Gyr.  From this, it is clear that younger stars tend to have lower \sig\ than older stars, but
there are complexities; the history of the merger has been imprinted onto the dynamics of the remnant. The stars
that formed before the first pass (i.e., the stars  older than $3.0$~Gyr) have had time to become dynamically heated. As
we saw in Figure~\ref{fig6}, these stars were initially rapidly heated during the first pass and then gradually heated
during the remainder of the merger. Stars that formed immediately after the first pass have had fewer opportunities to
become mixed and heated. As mentioned in the discussion of Figure~\ref{fig6}, many of these stars formed in tidal dwarf
galaxies that underwent a decrease in \sig\ after the second pass. Consequently, the oldest of these stars have the
lowest value of \sig, so the slope of the relation is inverted for stars between 1.7~Gyr and 2.7~Gyr old. Stars that
formed during and after the second pass have had even fewer opportunities to become dynamically heated. All of the
stars that were born during the second starburst (indicated by the vertical dashed line labeled \textit{g}) formed
either in the nuclear cluster of the newly-coalesced system or the satellite galaxies, with the majority forming in
the nuclear cluster. These stars cooled dynamically over time as the orbits of the satellite galaxies decayed. Finally,
the majority (85\%) of the stars with ages less than 0.5~Gyr, formed in nuclear disks with very low velocity dispersion
and have not had time to become substantially heated.

\section{ADDITIONAL STATISTICS}

\subsection{AGN Activity}

Observations examining the cosmic evolution of the \msig\ relation
\citep[e.g.,][]{treu2004,treu2007,woo2006,woo2008,hiner2012,canalizo2012} appear to indicate that \smbhs\ formed more
rapidly than their host galaxies; given a fixed value of \sig, galaxies at redshifts of $z>0.1$ have
more massive black holes than local galaxies. Unfortunately, in order to measure \mbh\ in non-local galaxies, a \smbh\
must be actively accreting gas. Exclusively using AGN host galaxies in such studies raises the concern that the sample
may be biased in various ways. For example, AGNs are often associated with galaxy merger activity
\citep[e.g.,][and references therein]{canalizo2013}.
Depending on the timing of the AGN activity with respect to the merger activity, measuring \sig\ in AGN hosts galaxies
could introduce extra scatter in the resulting \msig\ relation or it could systematically bias the value of \sig\ to
higher or lower values, leading to an artificial offset.

In order to determine whether \sig\ differs statistically between AGN host galaxies and inactive galaxies, we
examined the dynamical circumstances under which significant accretion occurred during each of our simulations. The
characteristic \smbh\ accretion timescale in our simulations was shorter than, or comparable to, our resolution limit
of 5~Myr; the accretion rate frequently changed by factors of 10--100 between consecutive snapshots. Consequently,
we likely did not capture \textit{all} of the enhanced accretion activity. Nevertheless, by examining all of our
simulations, we were able to clearly identify periods during which significant accretion was likely to occur as well as
periods during which significant accretion was not likely. For a detailed discussion of AGN lifetimes in hydrodynamic
simulations similar to ours, as well as a summary of observational evidence, see \citet{hopkins2006},
\citet{hopkins2009} and references therein. We found that accretion corresponding to a bolometric luminosity of
$10^{44}\,\ergps$ appeared to be a natural threshold separating the most luminous AGN activity from the much more
frequent periods of less significant accretion. Incidentally, $10^{44}\,\ergps$ is also commonly adopted as the
threshold separating quasars and Seyfert galaxies, so we use the phrases ``significant accretion'' and ``quasar-level
accretion'' interchangeably.

\begin{figure}
    \includegraphics[width=3.3in]{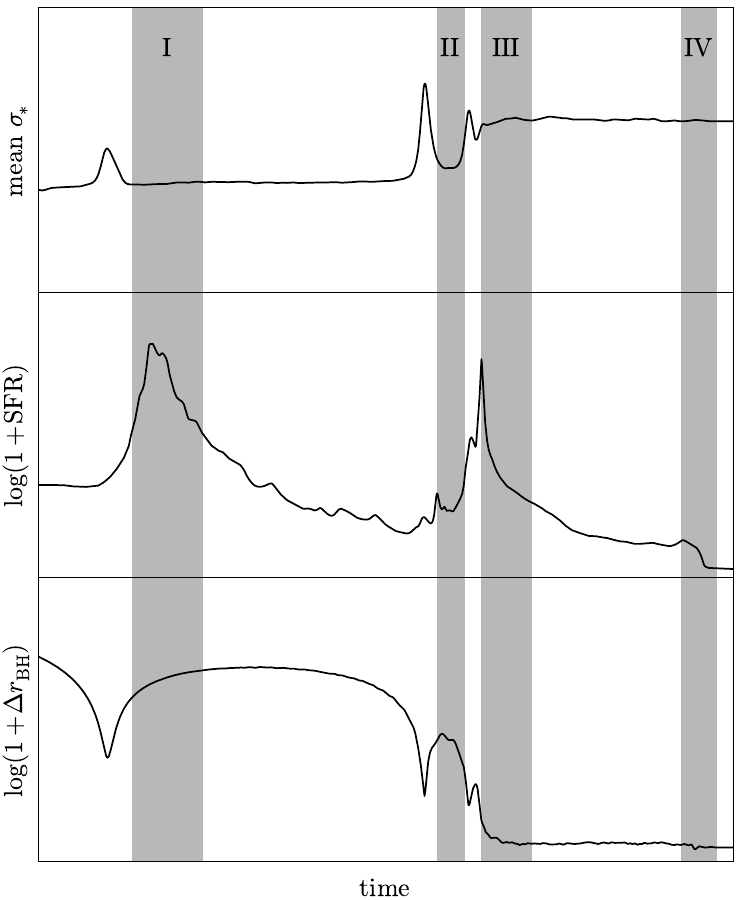}
    \caption{\label{fig8}\footnotesize Periods of significant accretion during a generic merger.
    All significant accretion (with $L_{\rm bol}\gtrsim10^{44}\,\ergps$) occurred during one of four periods: (I)
    Shortly after the first pass, (II) between the second and third passes, (III) during and immediately following
    nuclear coalescence, and (IV) long after nuclear coalescence. Not all mergers exhibited significant \smbh\
    accretion during all four periods. }
\end{figure}

All significant (quasar-level) accretion occurred during four periods. In Figure~\ref{fig8}, these periods are shown
with gray shading along with generic merger time series plots of \sig, star formation rate, and black hole separation.
Period I occurred shortly after the first pass while the progenitors were receding from one another. Period II occurred
between the second and third passes. Both progenitors hosted quasars during these periods, but usually not
simultaneously; the quasars turned on and off independently of one another, as discussed in more detail by
\citet{vanwassenhove2012}. This suggests that binary quasars with separations of 10--100~kpc are rare, relative to the
occurrence of quasars in general. Period III began at the moment of nuclear coalescence and period IV occurred long
after coalescence, when some of the gas that was not significantly heated during period III fell toward the nucleus.
Periods III and IV were associated with the most luminous quasars observed during our mergers, with $L_{\rm
bol}\sim10^{45}\,\ergps$. Interestingly, quasar-level accretion was never observed during the second or third passes
when the velocity dispersion was substantially elevated. This may be due to the $v_\infty$ term in the denominator of
Equation~(\ref{eq1}), since the relative speed of the \smbh\ with respect to the surrounding gas tends to increase
during the collisions. Accretion corresponding to bolometric luminosities of $L_{\rm bol}<10^{44}\,\ergps$ occurred
sporadically at all times after the first pass---including the second and third passes. We note that quasar-level
accretion did not occur during all four periods in each simulation, however when quasar-level accretion was detected, it
was always during one (or more) of the four periods identified in Figure~\ref{fig8}. While this does not imply that
quasar-level accretion never happens during other stages of the merger, it suggests that quasar activity is rare during
other stages of merger evolution.

For each simulation, the mean offset of \sig\ from the fiducial value was computed during each of the four quasar
periods. There was no detectable offset in \sig\ during periods I, III, and IV, however an offset was always
present during period II. Since the fiducial value of \sig\ during period II is somewhat ambiguous, we computed two mean
fractional offsets: the offset from the progenitor system, $(\sigmath-\sigma_{\rm prog})/\sigma_{\rm prog} =
0.11\pm0.05$, and the offset from the final remnant system, $(\sigmath-\sigma_{\rm final})/\sigma_{\rm final} =
-0.28\pm0.02$. Due to these offsets, the inclusion of period II quasar host galaxies in an observational sample
could potentially introduce extra scatter, or an offset, in a plot of the \msig\ relation. In Figure~\ref{fig8}, we see
that the \smbhs\ are significantly separated during period II. Therefore, a measurement of \mbh\ in such a system would
correspond with the mass of a \textit{progenitor} \smbh. The fiducial value of \sig\ used in the \msig\ relation
would then be the \textit{pre-merger} value of the progenitor spheroid. If we assume that progenitor systems generally
fall within the scatter of the local \msig\ relation, then observations of period II systems would tend to have high
values of \sig\ relative to their \mbh. Stated differently, these systems would appear to have under-massive black holes
when placed on the \msig\ diagram. Thus, the overly massive black holes that are observed at high redshift cannot be
due to measurements of period II systems. Furthermore, in our simulations, period II quasar activity accounted for only
16.1\% of all quasar-level AGN activity; it is unlikely that a large fraction of randomly selected quasar hosts would
consist of period II systems. Finally, from the images of the period II system (snapshot~\textit{e}) in
Figure~\ref{fig2}, it is evident that period II systems are composed of two distinguishable galaxies (when viewed along
most lines of sight), so they should be relatively easy to identify.

When interpreting these results, one should be aware that the timing of quasar activity depends upon
the treatment of hydrodynamics and \smbh\ feedback in our simulations. We have tried to make our results more robust by
considering only the general periods of likely accretion, rather than the exact timing of the accretion.
However, recent work by \citet{hayward2013} suggests that the hydrodynamic evolution in the late stages of \gadget\
simulations can differ significantly from the evolution observed in more realistic simulations when \smbh\
feedback is included. This casts some doubt on the timing and prevalence of Period IV quasars, but our general finding
remains unchanged; during a period of quasar activity, \sig\ is not likely to be strongly offset from its fiducial
value. Even if Period IV quasars \textit{never} occur in nature, the majority of quasar activity still occurs during
periods when \sig\ is not significantly offset. Alternatively, if Period IV accretion is more likely in reality than our
simulations suggest, then observing a quasar host with an elevated velocity dispersion would be more rare than our
simulations suggest, since Period IV quasars occur after the \sig\ has reached a stable value.

\subsection{Intrinsic Scatter \label{scatter}}

Measurements of \sig\ in real galaxies are necessarily made from random viewing directions at random times during
galactic evolution. There is no way of observationally determining the intrinsic scatter of \sig\ with respect to its
quiescent, fiducial value. Using our simulation data, we are able to provide estimates of this intrinsic scatter. Since
observed systems can broadly be categorized as either ongoing mergers or passively evolving (or simply ``passive'')
systems, we present two intrinsic scatter estimates---one for passive systems and one for ongoing mergers. In this
analysis, three conditions must be met for a galaxy to be considered passive:

\begin{enumerate}
    \item The galaxy must be clearly distinguishable from neighboring galaxies.
    \item The galaxy must contain only one nucleus.
    \item The galaxy must contain at most one large disk structure.
\end{enumerate}

\noindent
If any of these general criteria are not met, then the system is considered an ongoing merger. Passive systems
include all systems that appear to be non-interacting as well as systems that have clearly undergone recent
interactions. For example, even though the progenitors in simulation S1 show strong signs of interaction after the
first pass (see snapshot~\textit{b} in Figure~\ref{fig2}), we classify them as passive galaxies between approximately
0.5~Gyr and 2.0~Gyr. The system is also classified as passive immediately after nuclear coalescence, at 2.4~Gyr, even
though there are signs of recent interaction, such as stellar shells (see snapshots~\textit{g} and \textit{h}). We
classify simulation S1 as an ongoing merger between 2.0~Gyr and 2.4~Gyr (snapshots~\textit{c}--\textit{f}) and also
during the first pass, between approximately 0.25~Gyr and 0.4~Gyr. Admittedly, there are special circumstances that can
cause an ongoing merger to \textit{appear} to be a passive system and vice versa. In the present paper, we
have ignored these effects.

\begin{figure}
    \includegraphics[width=3.3in]{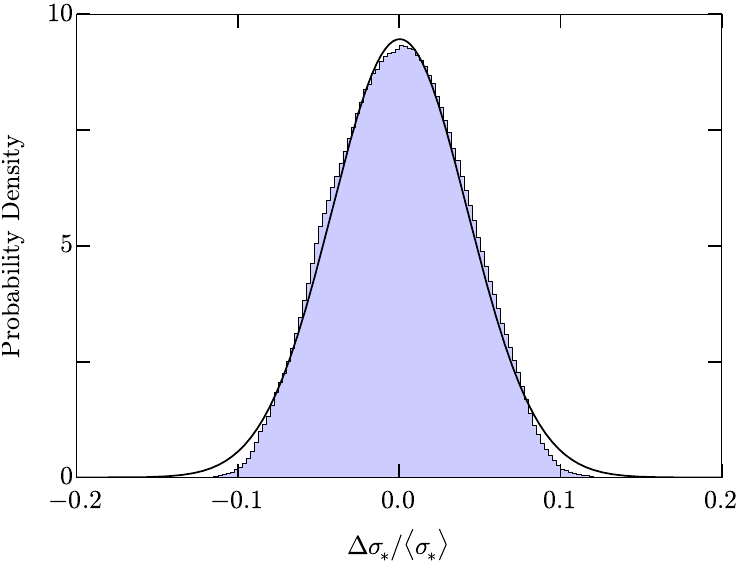}
    \caption{\label{fig9}\footnotesize Scatter probability distribution for coalesced systems, where $\Delta\sigmath$
    is the offset from the mean value of stellar velocity dispersion, $\langle\sigmath\rangle$. This plot includes
    data from all snapshots from simulations S1--S8) saved during periods of passive evolution, as defined in
section~\ref{scatter}. The best-fitting Gaussian, with $\sigma = 0.042$ and $\mu=1.25\times10^{-5}$ is over-plotted.
}
\end{figure}

\begin{figure}
    \includegraphics[width=3.3in]{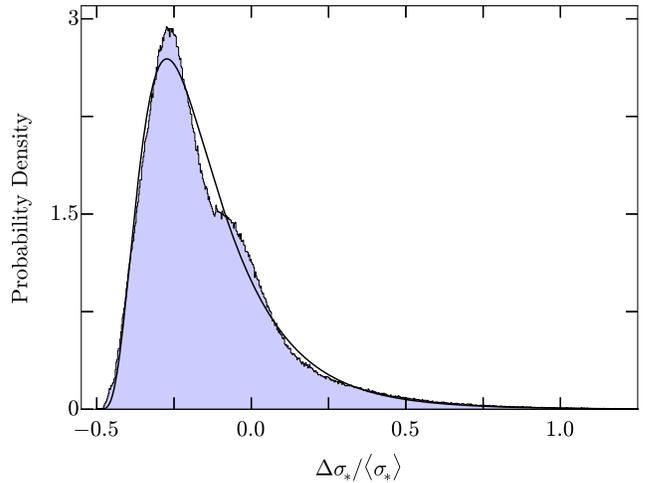}
    \caption{\label{fig10}\footnotesize Scatter probability distribution for merging systems, where $\Delta\sigmath$
    is the offset from the mean value of stellar velocity dispersion of the final remnant,
    $\langle\sigmath\rangle=\sigma_{\rm final}$. This plot includes data from all snapshots from simulations S1--S8
saved between the onset of the second pass and nuclear coalescence. The best-fitting log-normal distribution, with
$\sigma = 0.543$, $\mu=-1.150$, and $\delta=-0.509$, is over-plotted.}
\end{figure}

Upon separating each merger simulation into periods of ongoing merger activity and periods of passive evolution, we
computed the probability distribution of the fractional offset, $\Delta\sigmath/\langle\sigmath\rangle$ from the
fiducial value. The quantity $\Delta\sigmath = \sigmath - \langle\sigmath\rangle$ is the offset from the current
fiducial value, $\langle\sigmath\rangle$. In the passive period after the first pass, the fiducial value is the time
average of the mean \sig\ time series during that period. In all other cases, the fiducial value is the time average of
the mean \sig\ time series in the remnant system, $\sigma_{\rm final}$. In Figure~\ref{fig9}, we present the probability
distribution for passive systems. This plot contains data from all of our simulations. The best fitting elementary
distribution (in the least squares sense) was the Gaussian,

\begin{equation}
\frac{dP}{ds} = \frac{1}{\sigma\sqrt{2\pi}}\exp\left[{-\frac{(s-\mu)^2}{2\sigma^2}}\right]
\label{eq5}
\end{equation}
\noindent
with $\sigma=0.042$ and $\mu=1.25\times10^{-5}$. The corresponding plot for ongoing mergers is presented in
Figure~\ref{fig10}. The best fitting elementary distribution in this case was the shifted log-normal distribution,

\begin{equation}
\frac{dP}{ds} = \frac{1}{ \sigma\sqrt{2\pi}(s -\delta) }\exp\left[{-\frac{[\ln (s-\delta)-\mu]^2}{2\sigma^2}}\right]
\label{eq6}
\end{equation}

\noindent
with $\sigma = 0.543$, $\mu=-1.150$, and $\delta=-0.509$. In both cases, the distribution is more closely fit
by a linear combination of Gaussians; the above approximations are presented for simplicity. Using these densities, we
can easily compute various probabilities. For example, in the absence of measurement error, the probabilities of
measuring \sig\ within 5\% of the fiducial value (i.e., $\sigma_{\rm final}$) are, respectively, 0.77 and 0.10 for
passive systems and ongoing mergers. Furthermore, the probability of measuring \sig\ lower than the fiducial value
($\sigma_{\rm final}$) in an ongoing merger is 0.81.

While our approximation for the intrinsic scatter in passively evolving systems is likely fairly robust, the
approximation for ongoing mergers likely depends more heavily upon our merger parameters. Given the characteristic
directional distribution of \sig\ during a merger (see Figure~\ref{fig4}) and the temporal evolution (see
Figure~\ref{fig1}), it is clear that the distribution is strongly skewed, like the log-normal distribution presented
above. However, a greater variety of mergers would need to be examined in order to confidently compute the parameters
of the distribution in ongoing mergers.

\section{DISCUSSION AND CONCLUSIONS}

\noindent
In this paper, we have expanded upon the work presented in our previous paper (SC1) by examining the evolution of
stellar velocity dispersion in a suite of eight binary disk galaxy merger simulations that included dissipation, dark
matter, star formation, and AGN feedback. The analysis was designed, in part, to provide insight into observations of
\sig\ in systems that show signs of recent or ongoing merger activity. Our primary findings are as follows:

\begin{enumerate}
 \item During each merger, before the galactic nuclei coalesced, \sig\ underwent large, damped oscillations of
    increasing frequency. Once the nuclei coalesced, a series of small, statistically significant fluctuations continued
    until the remnant system became sufficiently mixed.  Qualitatively, this behavior is consistent with the
    findings of SC1, which examined the evolution of \sig\ in more idealized mergers of spherically symmetric,
    dissipationless systems that did not contain a separate dark matter component.

    \item Varying the gas fraction, and orbital parameters had no effect on the characteristic shape of the \sig\
evolution    time series. The level of apparent noise in the time series depended upon the gas fraction in a
non-trivial    way. Systems with larger gas fractions tended to spawn more tidal dwarf systems. These dwarf systems
added
    noise to the \sig\ time series because, while passing through the central region of their parent galaxy, they
caused \sig\    to briefly increase. However, the presence of dissipation and star formation evidently caused the
    phase-mixing stage of the merger process to be less pronounced; \sig\ oscillated more significantly during the
    phase-mixing stage of the dissipationless merger than in any of the dissipative mergers.

    \item No clear dependence was observed between the final velocity dispersion of the remnant and the gas
    fraction of the progenitors. However, $\sigma_{\rm final}$ was clearly lower in the completely
    dissipationless merger (S0) than in its dissipative counterparts (S1, S6, and S7). This is not surprising, since
    the dissipative merger remnants contained nuclear clusters of stars that formed during the simulation. The
    presence of a nuclear cluster deepened the gravitational potential well, leading to a larger \sig. In the
    absence of AGN and stellar feedback, even more mass would have likely accumulated in the nuclear region, causing
    $\sigma_{\rm final}$ to be even higher. We could not test this quantitatively because running simulations without
feedback was    prohibitively expensive, due to the formation of many dense clumps of gas and stars.

    \item Mergers of larger mass ratio (i.e., major mergers) exhibited the most significant absolute fluctuations in
    \sig. However, the \textit{relative} size of the fluctuations was not sensitive to the mass ratio. To see this,
    refer to the data from simulations S1, S4, and S5 in Table~\ref{table3}. These were, respectively, 1:1,
    1:2, and 1:4 mergers with otherwise identical initial parameters. The value of \sig\ at the climax of the
    second pass ($\sigma_2$), was highest in S1 and lowest in S5, however, there was no trend in the
    fractional increase, $\sigma_2/\sigma_{\rm final}$.

    \item When \sig\ is measured in systems that contain two progenitors moving relative to one another along the line
    of sight, the resulting measurements are artificially elevated because the streaming motion of the progenitors is
    mistaken for velocity dispersion. Equation~\ref{eq3} relates the apparent velocity dispersion of the combined
    system with the relative line-of-sight velocity and instrinsic properties of the progenitor systems.

    \item During galaxy collisions, \sig\ increases in all directions. The enhancement in \sig\ is greatest
    along the collision axis, partially because the bulk motion of the two progenitor systems can be mistaken for true
    velocity dispersion, as noted above. Conversely, the enhancement is lowest along lines of sight perpendicular to
    the collision axis. The mean of \sig\ over the set of all possible viewing directions closely traces the intrinsic
    velocity dispersion of the system.

    \item Stars in our simulations were born with lower \sig\ than that of the system as a whole. The apparent velocity
    dispersion of the youngest stars in the nuclear disks of our remnant systems was lower than the global stellar
    velocity dispersion by an average of $\approx30~\rm km~s^{-1}$. New stars tended to become
    dynamically heated with time unless they were tightly bound into clusters or dwarf galaxies. The velocity
    dispersion of stars residing in dwarf galaxies \textit{decreased} with respect to the global system as the orbits
    of their parent systems decayed due to dynamical friction.

    \item Quasar-level accretion activity was not detected during times when \sig\ was strongly enhanced. On the
    other hand, Seyfert-level accretion occurred sporadically at all times after the first pass. In general, AGN
    activity does not preferentially occur when \sig\ is strongly offset from its fiducial, equilibrium value. This is
    consistent with recent observational evidence \citep{woo2013}, indicating that active galaxies fall on the same
    \msig\ relation as quiescent galaxies.

\end{enumerate}

Given our findings, we would advise anyone who is interested in measuring \sig\ in a dynamically questionable system
to note the following:

\begin{itemize}
    \item When \sig\ is measured in systems which clearly contain two nuclei, the resulting value of \sig\ in the
individual nuclei depends upon the nuclear separation distance. Nuclei that are significantly separated (e.g.,
snapshot~\textit{e} in Figure~\ref{fig2}) are likely to retain a value of \sig\ that is only slightly elevated with
respect to the pre-collision value of the progenitor (compare the histograms for progenitor~A and snapshot~\textit{e} in
Figure~\ref{fig4}). As the distance between the nuclei decreases, \sig\ increases. When the nuclei are strongly
superposed, as in snapshot~\textit{c}, the measured value of \sig\ is likely to be higher than the value of \sig\ in the
eventual remnant system. Of course, projection can also cause significantly separated nuclei to appear to be
significantly superposed, so significant nuclear superposition is a weak diagnostic.

    \item A measurement of \sig\ is likely to be elevated relative to the eventual remnant value if the system
     contains two or more disk-like structures, but only one visible nucleus, as seen in snapshot~\textit{c}
     of Figure~\ref{fig2} (viewed along the $\theta=75^\circ,\,\phi=66^\circ$ direction).

    \item Measurements of \sig\ in systems that contain only one nucleus are likely robust if the system also contains
    stellar shells or exhibits a dynamically relaxed morphology (see snapshot~\textit{h}). Shells tend to form
    \textit{after} \sig\ has reached its stable post-merger value. More generally, if a system appears to be
    dynamically relaxed, a measurement of \sig\ is likely robust; the presence of low-surface brightness debris in
    the region surrounding a galaxy that otherwise appears to be relaxed does not indicate that \sig\ is enhanced.

    \item Measurements of \sig\ in the bulge components of disk-like systems containing strong bridges or tidal
    tails (e.g., snapshot~\textit{b}) are not likely to differ from the value of \sig\ measured in the bulge before the
    interaction took place.

    \item Systems with quasar-level luminosities ($L_{\rm bol}\gtrsim10^{44}\,\ergps$) are unlikely to have
    substantially elevated or suppressed values of \sig, relative to the fiducial, equilibrium value.

\end{itemize}

For a concrete example, consider the prototypical ongoing merger, NGC 6240. This system contains two nuclei with a
projected separation of $\sim800$~pc. Using the guidelines outlined above, we would expect the velocity dispersion of
each progenitor nucleus to be mildly elevated with respect to its pre-merger value since two nuclei are visible, but
they are not separated by a large distance. \citet{medling2011} measured \sig\ and \mbh\ in the southern nucleus of NGC
6240 and found that the nucleus lies within the scatter of the \msig\ relation. Assuming that (1) the nucleus was on the
relation before the merger began and (2) the \smbh\ has not grown substantially since the beginning of the merger, then
this finding is consistent with what we expect. However, note that the measurement of $\sigma_*=282\pm20 \kmps$ by
\citet{medling2011} was based upon the dynamics of the CO bandheads of later-type giants and supergiants within 300~pc
of the southern black hole, so the measurement may be lowered due to the presence of a dynamically cool nuclear disk.
Also, if NGC 6240 were placed at sufficiently high redshift, or the observations were of lower resolution, the two
nuclei would not have been distinguishable. In this situation, we would only be able to classify NGC 6240 as a generic
ongoing merger. Based upon the scatter analysis of Section~\ref{scatter}, we see that a measurement of \sig\ in such a
system is 81\% likely to be lower than the value of \sig\ in the relaxed remnant (with the most likely measurement being
27\% lower than the final value). \citet{oliva1999} measured \sig\ of the entire merging system using the Si 1.59
$\mu$m, CO 1.62 $\mu$m, and CO 2.29 $\mu$m lines, obtaining measurements of 313 \kmpst, 298 \kmpst, and 288 \kmpst,
respectively. These measurements of \sig\ place the southern black hole, together with the \sig\ of system as a whole,
within scatter of the local \msig\ relation. Once the two progenitor \smbhs\ merge, we would expect \sig\ to increase in
order for the system to remain on the \msig\ relation; this is consistent with our expectation that a measurement of
\sig\ in NGC 6240 is likely to be lower than that of the eventual remnant.

The reader should be aware that the conclusions above were based upon a fairly small number of simulations which
were performed using an imperfect simulation code. While several initial conditions were independently varied, extreme
cases were not tested. The simulations did not have sufficient resolution to follow the evolution of individual stars or
the detailed structure of the multi-phase interstellar medium. It should also be noted that the recipes used for star
formation and \smbh\ feedback were very crude and cannot be expected to faithfully represent reality. Furthermore, all
measurements of \sig\ in this paper were mass-weighted. In order for this work to be more relevant to observational
studies, we need to know whether mass-weighted determinations of \sig\ are consistent with the flux-weighted
measurements that are performed during observations of real galaxies. In SC1, we showed that, in principle,
flux-weighted \sig\ can differ from mass-weighted \sig\ in the presence of dust extinction. In the simulations of the
present paper, we have seen that the intrinsically more luminous new star particles tend to be dynamically cooler than
the older population of less luminous particles. In the next phase of this research (N. R. Stickley et al., in
preparation), we plan to characterize potential differences between flux-weighted and mass-weighted determinations of
\sig\ by creating synthetic Doppler-broadened spectra, generated using the kinematics feature of the radiative transfer
code, \textsc{Sunrise} \citep{jonsson2006,jonsson2010}. This will allow us to characterize the effect of dust
attenuation on measurements of \sig\ in a much more realistic manner than previously done in SC1.

\acknowledgements

We thank Gillian Wilson and Desika Narayanan for their invaluable assistance, Jong-Hak Woo for making helpful
suggestions, Barry Rothberg for his insights, and the referee, Joshua Barnes, for his helpful feedback. This
work used the Extreme Science and Engineering Discovery Environment (XSEDE), which is supported by National Science
Foundation grant number OCI-1053575. Financial support for this work was provided by NASA through a grant from the Space
Telescope Science Institute (Program numbers AR-12626 and GO-11557), which is operated by the Association of
Universities for Research in Astronomy, Incorporated, under NASA contract NAS5-26555. Additional support was provided by
the National Science Foundation, under grant number AST-0507450.

\end{document}